# Formation of bound dineutrons in the $^{175}$Lu (*n*, $^{2}n$) $^{174g}$Lu nuclear reaction and its cross section

Ihor Kadenko [a, b, *], Barna Biró [c], Mihály Braun [c], András Fenyvesi [c], Kateryna Okopna [b], Nadiia Sakhno [a, b], Loránd Zakàny [d]

[a] *International Nuclear Safety Center of Ukraine of Taras Shevchenko National University of Kyiv, 01601, Kyiv, Ukraine*
[b] *Department of Nuclear and High Energy Physics, Faculty of Physics, Taras Shevchenko National University of Kyiv, 01601, Kyiv, Ukraine*
[c] *HUN-REN Institute for Nuclear Research (ATOMKI), Debrecen, Hungary*
[d] *Isotoptech Zrt., Debrecen, Hungary*

**Abstract**

The dineutron as a bound chargeless nucleus of two identical nucleons has been attracting attention for many decades. In addition to the study of the formation of a bound dineutron in the outgoing channel of specific nuclear reactions, cross section estimates have been gaining interest as research targets. We used a traditional neutron activation technique to irradiate Lu samples followed by measurements of the induced activity with HPGe spectrometer in order to detect gamma-peaks of the $^{175}$Lu (*n*, $^{2}n$)$^{174g}$Lu reaction product when this reaction channel is not open for incident neutron energies. Based on two measurements of the instrumental gamma ray spectra of Lu samples we reliably identified the presence of the $^{174g}$Lu isotope in the outgoing channel of the $^{175}$Lu (*n*, $^{2}n$)$^{174g}$Lu nuclear reaction. Also, the cross section for this nuclear reaction was obtained and to be equal 33.5±5.9 mb. For the first time, the dineutron as a nuclear reaction product was indirectly detected via statistically significant observation of the induced activity of $^{174g}$Lu with corresponding cross section estimate.

**Keywords:** Bound dineutron, neutron induced nuclear reaction, Lutetium-175 isotope, reaction cross section, binding energy interval estimate

## 1 Introduction

The study of a bound system of two nucleons when both of them have no charge is the best tool to deeply understand the nature of the strong interaction. Moreover, N=2 is the very first magic number and due to this fact the dineutron is expected to exist for a certain long lifetime. Therefore, search for a bound dineutron as a charge-symmetric partner of the deuteron has been ongoing for many years, though many experiments, including the latest ones, came up empty [1-3]. Worth noting that the subject of this letter does not overlap with "dineutron correlations", by definition being a spatial correlation of two valence neutrons localized at similar positions inside light neutron excess nuclei [4]. At the same time, the theoretical substantiation of the existence of two identical-nucleon bound systems in the potential well of a heavy nucleus was discussed in [5] and detailed in [6]. In these works, a quantum mechanical problem is solved for a system of two interacting particles in the potential well. From the solution of this problem (the aforementioned problem, named) it follows that near the surface of a heavy nucleus and within its potential well (0-0.4 MeV deep into the well), there are discrete energy levels at which two particles can be in a bound state, thus forming a system, similar to the lightest nuclei. The simplest representative of such two identical nucleon systems is the dineutron in a bound state. Based on the solution of the problem in [5, 6], in the article [7] a set of medium and heavy weight nuclei (90<A<210) was ranked according to the criteria that characterize the likelihood of dineutron formation. For example, $^{159}$Tb and $^{197}$Au as target nuclei are included in the "most-likely" category. The results of the corresponding experiments are covered in [8, 9] and confirmed these delayed conclusions, paving a reliable way for successful research of such a unique bound system as the dineutron is. Isotope nucleus $^{174}$Lu, according to this ranking, is in the "less-likely" category because only two criteria out of four are met. At first glance, it would be worthwhile to perform the experiments with other much better promising candidate nuclei. However, knowledge of boundaries, within which the dineutron can be observed with necessary statistical significance, is also very important. Thus, it was decided to conduct an experiment in order to study whether it could be possible to observe the formation of bound dineutrons under conditions that are not the most favorable, and to determine the cross section for this reaction. The results of this experiment are covered and discussed in this letter.

## 2 Experimental setup

### 2.1 Neutron irradiation facility

During the experiment the neutron activation method was utilized with application of the MGC-20E cyclotron. The Lu sample was exposed with quasi-monoenergetic *d*+D neutrons, generated using the $^{2}$H(*d*, *n*)$^{3}$He nuclear reaction. Deuterons, accelerated to energy $E_d$ = 4.354 MeV ± 0.4%, were transported to the irradiation site. Details of beam collimation [10] and sketch of the irradiation are shown in Fig. 1. The beam current measured on the neutron emitting target was $I_d$ = 1.82 μA ± 3%. A gas cell filled with $D_2$-gas (height - 39.6 mm, diameter - 40 mm) was used as a target for deuteron irradiation. The vacuum window of the gas cell is covered with 25 μm Nb foil. The pressure in the chamber was kept at the level of $<p_{gas}>$ = 2.22 bar ± 13%, the average loss of the target pressure was $<dp/dt>$=0.04 bar/h. The Lu sample consisted of two disks made of the same naturally isotopic Lu material. Both disks had a diameter of 20 mm and a height of 1.2 mm. The masses of the disks are 6.387 g and 6.295 g. The content of rare earth elements in the material was checked using microwave plasma atomic emission spectroscopy (MP-AES). The results obtained for the elemental composition of the two samples are as follows: $^{nat}$Lu - 96.45 w%; Hf - 0.05 w%; Th, Pm, Gd, Tb, Dy and Yb are present, but no quantitative analysis was done.

* Corresponding author.
E-mail address: imkadenko@univ.kiev.ua

With application of the NeuSDesc code [11] the energy of the neutrons emitted was in the $E_n$ = [5.51÷6.00] MeV energy range with $<E_n>$ = 5.76 MeV mean energy in the volume of the Lu sample. The estimated volume averaged neutron fluence rate was $<\varphi_n>_{Id=1\mu A}$ = 3.4×10$^6$ cm$^{-2}$s$^{-1}$μA$^{-1}$ ± 17% for the Lu sample. The irradiation time was $t_{irr}$ = 56,640 s and the neutron fluence averaged for the volume of the Lu sample was $\Phi_n \cong 3.6\times10^{11}$ cm$^{-2}$.

*2.2 Detector setup*

The induced gamma activity of the neutron activated Lu sample was done at ATOMKI using a MIRION CANBERRA GX2018 type vertical XtRa HPGe detector equipped with a Model iPA-SL preamplifier used with negative output polarity. The energy resolutions of the detector are at FWHM = 0.850 keV at $E_\gamma$ = 122 keV and FWHM = 1.8 keV at $E_\gamma$ = 1.33 MeV gamma energies. Multichannel analyzer and the Genie™ 2000 Gamma Acquisition & Analysis (V3.4.1, Nov. 1, 2016, Mirion Technologies (Canberra), Inc.) software was employed for recording 8,192 channel gamma spectra. The instrumental gamma spectra measured for calibrated $^{22}$Na, $^{57}$Co, $^{60}$Co, $^{137}$Cs, and $^{241}$Am radioisotope sources were used for the energy calibration of the detector.

The HPGe detector and the Lu samples were in a shield with 10 cm thick walls made of lead. The lining inside the shield consists of a 1 mm thick copper layer and a 1 mm thick cadmium layer that is in between the copper layer and the lead wall. For counting the Lu samples were placed on a 1 mm thick plastic sheet made of polymethyl methacrylate (PMMA) and the cryostat window was covered with a cap with 3 mm thick walls made of polyethylene (PE). The distance between the upper surface of the carbon epoxy composite window of the cryostat and the lower surface of the Lu sample was 24 mm. Thus, the distance of the lower surface of the Lu sample and the front surface of the crystal was (63.0 ± 0.5) mm.

*2.3 Detection efficiency vs $E_\gamma$*

Considering a short distance between the Lu sample and the detector none of them could be treated as point-like objects. Therefore, the detection efficiency as a function of the gamma energy was calculated with the ISOCS™ (In Situ Object Counting System) Calibration Software (ISOCS™ is a trademark of Canberra Industries, Inc.).

First, the geometry model of the counting arrangement including the HPGe detector was constructed and then the relevant input file was prepared for the simulation. Then the detection efficiency was calculated and the roles of the cascade summing corrections were studied, too. It has been found that the cascade summing corrections are negligible compared to the statistical uncertainties of the net peak areas evaluated from the measured gamma spectra. Uncertainties of calculated efficiency values were up to 4.01%.

*2.4 Instrumental spectra acquired*

Four spectra were taken using the experimental setup described above in Table 1.

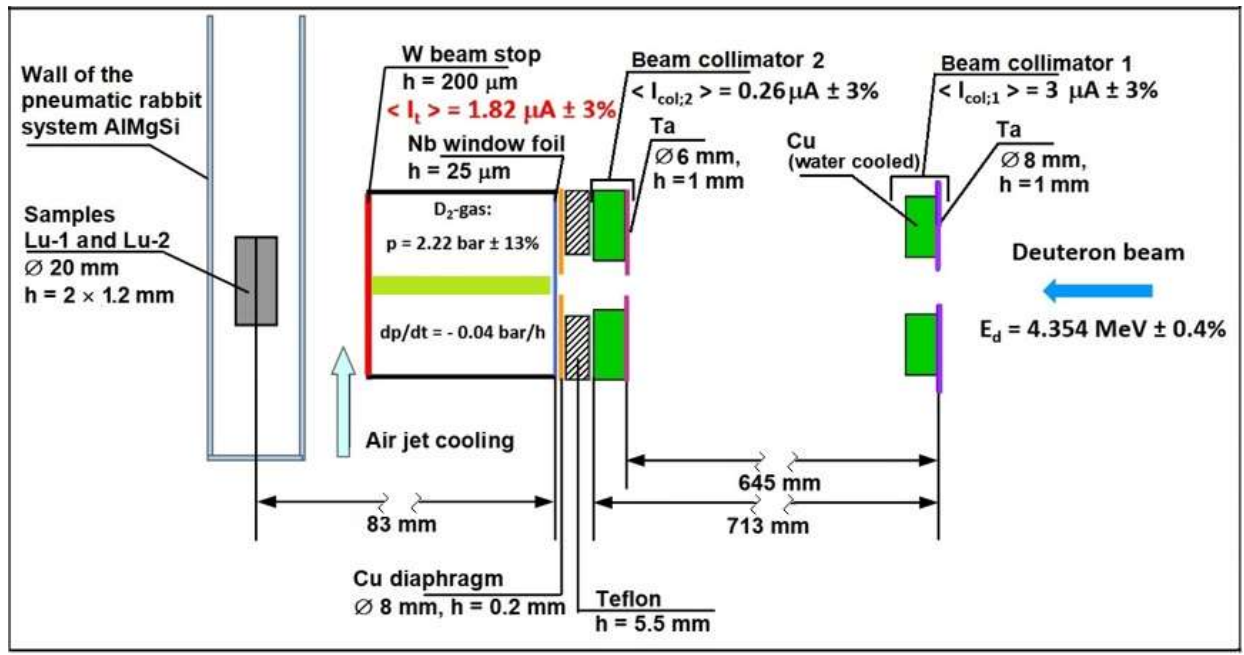

**Fig. 1.** The sketch of the irradiation arrangement.

**Table 1.**
Summary of the gamma counting measurements performed after the irradiation stopped at 2023.05.19 17:00:00. (Detector: MIRION CANBERRA HPGe XtRa).

| Measurement ID | START | Live time (s) | Real time (s) |
|---|---|---|---|
| Lu-1 +Lu-2 1st | 5/22/2023 7:23:00 AM | 713,893.3 | 715,246.1 |
| Lu-1 +Lu-2 2nd | 6/6/2023 10:25:55 AM | 596,883.6 | 597,972.3 |
| Lu-1 +Lu-2 3rd | 8/29/2023 3:17:00 PM | 1,887,971.7 | 1,891,309.3 |
| Background | 9/22/2023 5:34:00 PM | 402,231.1 | 402,596.1 |

We focused on data from two spectra: with the beginning of counting on May 22 and August 29.

## 3 Data analysis

### 3.1 Dineutron formation in the $^{175}Lu(n, {}^2n)^{174g}Lu$ nuclear reaction

Finally, in May-June and in August-September 2023 two spectra were acquired using the experimental setup described above: the spectrum 1 with the total acquisition 713,893.3 s live time s and 0.19% dead time; and the spectrum 2 with the total acquisition 1,887,971.7 s live time and 0.18% dead time. General view of the spectrum 2 is presented in Fig.2. As a result of the processing of the instrumental spectra, the following peaks with 76.5 and 1,241.8 keV energies were detected and presented in Fig. 3 below.

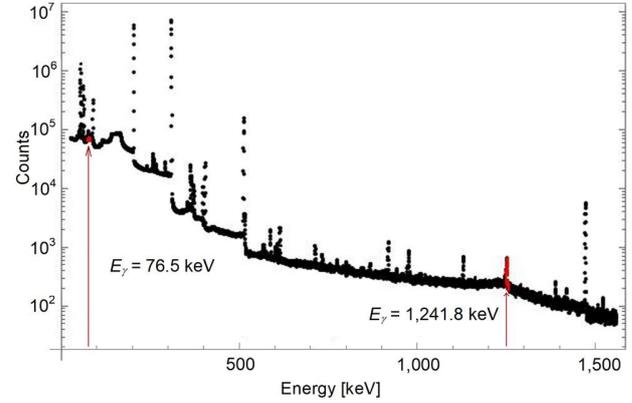

**Fig. 2.** General view of spectrum 2.

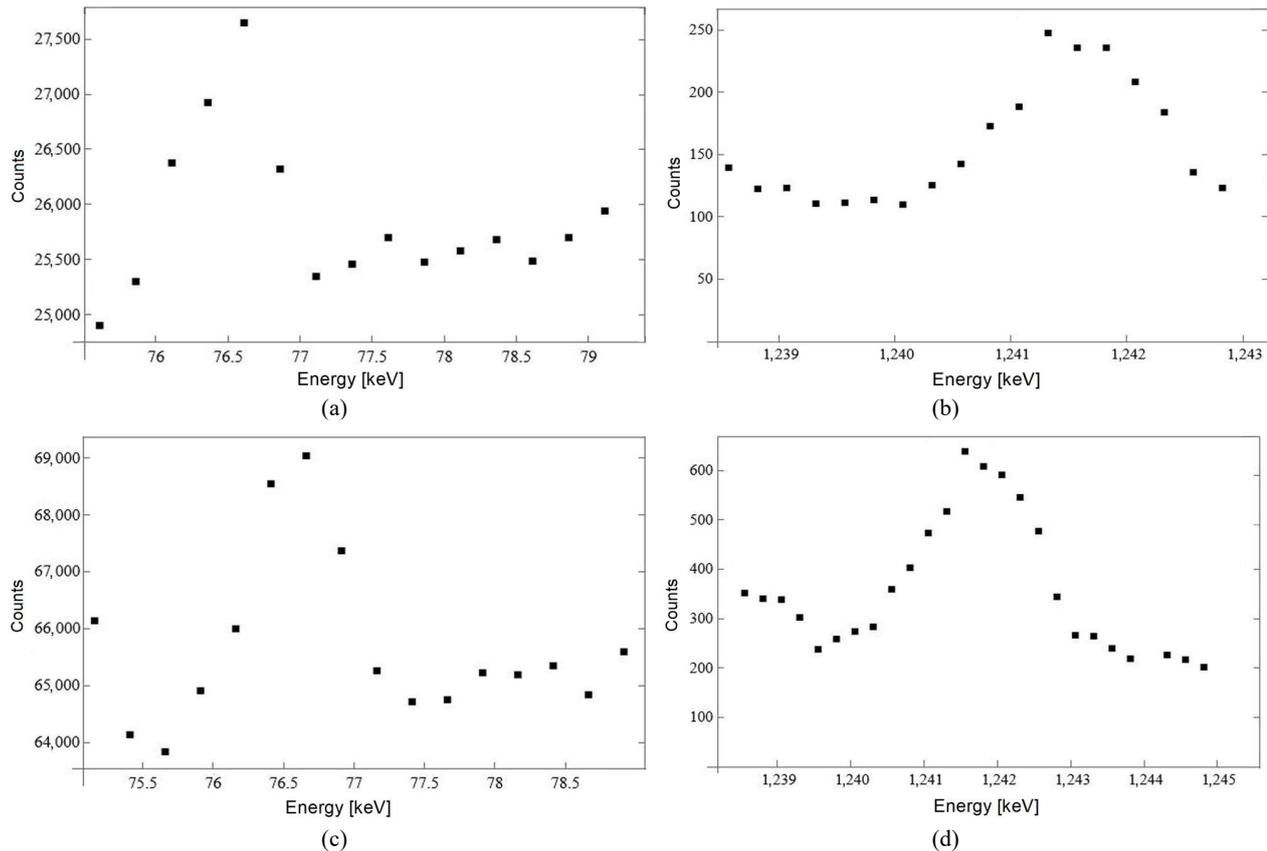

**Fig. 3.** Gamma peaks due to decay of $^{174g}Lu$ - spectra 1 and 2 with corresponding areas in counts: (a) $S_{1,a}$ = 5,883±370; (b) $S_{1,b}$ = 1,205±51; (c) $S_{2,c}$ = 13,610±601; (d) $S_{2,d}$ = 3,172±83 counts.

The background spectrum does not contain any peaks adjacent to 1,241.8 keV gamma-line due to $^{174g}Lu$ decay, but 76.5 keV energy peak is overlapping with $K\alpha$ series of Bi with 0.0028±0.0002 cps count rate out of total value of 0.0082±0.0005 cps with irradiated Lu samples present. Since $^{174m,g}Lu$ is not a natural isotope of lutetium and was not detected in the Lu samples before the irradiation, it is obvious that its formation is a consequence of reactions induced by the neutron flux directed at the sample. Let us consider the reactions that are possible on the $^{175, 176}$Lu or other impurity nuclei, taking as a basis the information about the energy of neutrons in the input channel as well as spectra data about gamma-line energies and intensities.

Thus, peak count rates for 1,241.8 keV energy from spectra 1 and 2: (169±7) ×10$^{-5}$ cps and (168±4)×10$^{-5}$ cps, correspondingly, are within uncertainties and this means that the half-life for this gamma emitter must be more than 30 months. In addition, both gamma-line energies must be within ±0.45 keV energy window and with comparable gamma-line intensities. Then according to information from the NuDat 3 database [12] the only candidate nucleus, suitable to meet all these 4 criteria for both gamma-lines is $^{174g}$Lu with the half-life 3.31 y. At this stage, we can preliminarily conclude that $^{174g}$Lu is detected with statistical significance far exceeding 5-sigma criterion in both instrumental gamma-ray spectra after irradiation of Lu samples of natural abundance with neutrons of energies, well below the energy threshold ($E_{th1}$=7.711 MeV) of the $^{175}$Lu(n, 2n)$^{174g}$Lu nuclear reaction. The last nuclear reaction could be a trivial explanation of $^{174g}$Lu appearance in the outgoing channel if (d, t) neutrons could be generated. There are three most likely reaction channels to produce tritium according to the design of our gas target (Fig.1): $^{nat}$W(d, t), D(d, n)t and $^{93}$Nb(d, t)$^{92m}$Nb. W beam stop and the D(d, n)t nuclear reaction were thoroughly investigated in [13] for deuterium beam energies 4 and 7 MeV with application of activation foils technique followed by unfolding the neutron field at 0° emission angle. In the investigated energy range, no self-target build-up was observed in the targets. In other words, no (d, t) or other greater energy neutrons were detected neither for 4 MeV nor for 7 MeV impinging deuterons. This intermediate result is proved by low cross section values (less than 10$^{-8}$ mb according to [14]) for $^{nat}$W(d, t) nuclear reactions with subsequent generation of (d, t) neutrons. Moreover, tungsten has very low solubility for hydrogen isotopes and is essentially unaffected by hydrogen accumulation for (d, t) neutrons generation. Tritium build-up in the D(d, n)t nuclear reaction was additionally considered in [9] and found to be for this research about 3,500 tritium atoms produced in the gas volume during all time of Lu samples irradiation. For the $^{93}$Nb(d, t)$^{92}$Nb nuclear reaction there is no experimental data in [15] and according to [14] we can expect a cross section evaluated value for tritium production at ~ 10$^{-10}$ mb for $E_d$ =4.354 MeV. Understanding, that the induced activity of $^{174g}$Lu is the product of these estimates, then about 90 mb of the (d, t) reaction to produce 14 MeV neutrons and about 2 b for the $^{175}$Lu(n, 2n)$^{174g}$Lu nuclear reaction, we can get the expected induced activity due to (d-t) neutron contribution from all target materials, including the gas volume, to be less than 0.2% of the total induced activity of both Lu samples.

Also, taking into account the results of the elemental and impurity MP-AES above as well as neutron field characteristics, the following interfering reactions were considered to be the potential contributors into the production of $^{174g}$Lu: $^{174}$Hf(n, p)$^{174g}$Lu; $^{176}$Hf(n, t)$^{174g}$Lu; $^{176}$Hf(n, n+d)$^{174g}$Lu. No one of these reactions is presented with the corresponding experimental data in the EXFOR database [15] for ~ 6.0 MeV neutron energy. In the same time, from the TENDL-2023 [14] database we can get the following cross section estimates for these reactions: 20 μb and 0 mb. For the $^{176}$Hf(n, n+d)$^{174g}$Lu nuclear reaction this reaction channel is not open for 6.0 MeV neutron energy. Then the nuclear reaction $^{174}$Hf(n, p)$^{174g}$Lu may be a potential contributor to 1,241.8 keV peak count rate. Before doing its estimate of a possible contribution, we considered another nuclear reaction: $^{174}$Hf(n, γ)$^{175}$Hf with 4.5 mb TENDL-2023 cross sections which is about 200+ times greater than 20 μb value. Therefore, the presence of 343.3 keV gamma-peak due to decay of $^{175}$Hf in the instrumental spectra could serve as an additional indicator for such a contribution of the $^{174}$Hf(n, p)$^{174g}$Lu nuclear reaction product. However, no such peak with 84% intensity was observed in our spectra and, hence, no reasonable contribution may take place to 1,241.8 keV gamma-peak due to the $^{174}$Hf(n, p)$^{174g}$Lu nuclear reaction.

In addition, both Lu samples were located behind the 2 mm AlMgSi wall of the pneumatic rabbit system to avoid even hypothetical direct hit reactions between Lu samples and charged particles like p, d, t, α as reaction products on target unit construction materials. Notwithstanding, the 2 mm AlMgSi wall could be the source of charge particles resulted in (n, x) reactions on Al, Mg and Si isotopes, therefore we checked the induced activity of Lu-2 sample, shielded by Lu-1 sample, and found it to be about a bit lower than a half of total induced activity of both samples. Finally, no significant contribution of neutrons above the threshold $E_{th1}$ to induce the observed activity of $^{174g}$Lu takes place and rather a low energy tail of neutron energy is expected according to [13] due to low enough energy of incident deuterons and neutron scattering effects. Our estimates are in full correspondence with the general conclusion in [16] that a deuteron beam and heavy target materials are featured with a very low neutron yield.

### 3.2 The cross section estimate for the $^{175}$Lu(n, 2n)$^{174g}$Lu nuclear reaction

Taking into account the features of 76.5 keV and 1,241.8 keV peaks formation in our instrumental spectra, we used only a gamma-line of 1,241.8 keV energy for cross section calculations. To estimate the $^{175}$Lu (n, 2n)$^{174g}$Lu nuclear reaction cross-section the following expression was utilized:

$$\sigma_{\exp} = \frac{N_{counts} \times \lambda \times A}{(1-\exp(-\lambda t_{irr})) \times \exp(-\lambda t_{cool}) \times (1-\exp(-\lambda t_{meas}))} \times \frac{1}{\xi \times I_\gamma \times \varphi \times N_A \times m \times p}, \quad (1)$$

where $N_{counts}$ - the number of counts under the full absorption peak 1,241.8 keV energy that accompanies the decay $^{174g}$Lu (1,205±51; 3,172±83 counts); $\lambda$ - nuclear decay constant of $^{174g}$Lu (6,64·10$^{-9}$ s$^{-1}$); A - mass number of lutetium atoms (174.967 u), $t_{irr}$ = 56,640 s, $t_{cool}$, $t_{meas}$ - duration of irradiation, cooling and measurement, respectively (s); $\xi$ - efficiency of γ-rays detection (0.01059±0.00042); $I_\gamma$ – gamma-quantum intensity due to the decay of $^{174g}$Lu nuclei (0.0514); $\varphi$ - flux density of fast neutrons ((6.19±1.05)×10$^6$ n/(s×cm$^2$)); $N_A$ - Avogadro's number (6.022×10$^{23}$ number of nuclei/mol); m – total mass of lutetium cylinders (6.387 g +6.295 g = 12.682g); p - absolute content of $^{175}$Lu isotope in lutetium targets

(0.9646×0.9741=0.9396).

The efficiency of detecting γ decay quanta was determined with the ISOCS Calibration Software (p. 1.3) taking into account geometrical dimensions of samples, self-absorptions of gamma-rays and cascade summing effects. The values $\lambda$, $p$, $I_\gamma$, $A$ were taken from [12], $N_{counts}$ - from the instrumental gamma-spectra.

Then using (1) the averaged estimate of the $^{175}$Lu $(n, 2n)^{174g}$Lu nuclear reaction cross section from two countings of Lu samples was obtained as follows:

$$\bar{\sigma}_{exp} = 33.5 \pm 5.9 \text{ mb}.$$

## 4 Discussion

In our earlier experiments and publications [8, 9] we have observed the formation of a bound dineutron in the $^{159}$Tb$(n, 2n)^{158g}$Tb and $^{197}$Au$(n, 2n)^{196g}$Au nuclear reactions. For the second reaction, the statistical significance of $^{196g}$Au decay was above five sigma. We were not the only researchers, who observed a similar effect that was not understood and explained earlier. Thus, in [17] the cross section of the $^{127}$I$(n, 3n)^{126}$I nuclear reaction was measured and equals 40±15 mb for impinging neutrons of 14.7 MeV energy, while the reaction threshold for this reaction is 16.42 MeV. This cross section value is comparable to our estimate above. In [18], authors claimed to obtain the $^{159}$Tb $(p, 3n)^{157}$Dy reaction cross section 90±10 μb for 14.86±0.85 MeV proton energy, which is below the energy threshold for this reaction: 17.14 MeV. No explications were given with respect to these results and the only reasonable explanation would be assumption about the formation of a bound dineutron in the outgoing channel. Then current results for the $^{175}$Lu$(n, 2n)^{174g}$Lu nuclear reaction extend the list of nuclei on which the formation of a bound dineutron takes place. Obviously, in our study the reaction $^{175,176}$Lu$(n, X)^{174g}$Lu, where $X$ is $2n$ or $3n$, takes place. The reaction thresholds in such cases are $E_{th1}$ mentioned above and $E_{th2}$=14,034.6 keV, respectively. The reaction with three neutrons separation is obviously impossible to occur, so let us focus on the separation of two neutrons. As mentioned above, the maximum neutron energy in the experiment was up to 6 MeV - so, if the $^{175}$Lu$(n, 2n)^{174g}$Lu reaction takes place, then it happens at a lower energy. The only thing that remains is to assume that bound two neutron states are observed in our experiment based on the energy conservation law: their formation requires less energy, so such sub threshold reactions are possible. Then the final result is as follows: a reaction takes place with the formation of a bound dineutron, namely the $^{175}$Lu$(n,2n)^{174g}$Lu reaction. Comparing the corresponding cross section estimates for the $^{159}$Tb $(n, 2n)^{158g}$Tb nuclear reaction [8] that equals 75±30 mb for 6.85 MeV neutron energy [19], and cross section values for the $^{197}$Au$(n,2n)^{196g}$Au nuclear reaction that were determined as 180±60 μb for the neutron energy range [6.09÷6.39] MeV and 37±8 μb for the neutron energy range [6.175÷6.455] MeV with current estimate 33.5 ± 5.9 mb for neutrons of $5.76^{+0.24}_{-0.25}$ MeV of the $^{175}$Lu$(n,2n)^{174g}$Lu nuclear reaction, one can conclude they are comparable for reactions on $^{159}$Tb and $^{175}$Lu (both belong to rare earth elements) and differ several times of those for $^{197}$Au. On one hand, this could mean that rare earth nuclei are more susceptible to bearing a bound dineutron than gold because of their much larger deformations. On the other hand, as it was mentioned in [5] by Migdal, these reactions must have a resonant structure of cross section vs energy dependence. Therefore, it is expected that for neutron beams of some energies with much less energy spread the corresponding cross sections can be much greater and for other energies – close to zero. Also, results of this experiment allow us to narrow an interval estimate for the binding energy of the dineutron ($B_{dn}$) and now instead of 1.3 MeV <$B_{dn}$< 2.8 MeV [7] we got a new range: 1.6 MeV <$B_{dn}$< 2.8 MeV. Being back to the case of light nuclei considered above, which are characterized by much lower two halo neutron separation energy (for instance, 369 keV for $^{11}$Li), our study supports the statement that no bound dineutron exists inside of $^{11}$Li [20].

## 5 Summary

In this letter, the formation of the dineutron as a bound state of two neutrons within the potential well and near (not at [4]) the nuclear surface of $^{174}$Lu in the reactions $^{175}$Lu $(n, 2n)^{174g}$Lu was observed for the first time. This fact is in full agreement with Migdal's prediction regarding the systems of such type to be located near the surface of some heavy nuclei at several fm distances beyond their volume but within the potential well. Therefore, our result does extend the list of possible candidate nuclei for generation of a bound dineutron, uniting now the blue, yellow and green groups of nuclei [7]. In addition, the cross section for this reaction was determined for the first time with application of a regular neutron activation technique. Thus, it is possible to draw conclusions about the course of the reaction with the formation of a dineutron near the $^{174}$Lu nucleus, despite the fact that, according to the corresponding ranking of nuclei, it is not best suited for this phenomenon. Further research and experiments are needed to deepen our understanding of this process from the point of view of fundamental physics and because this phenomenon may soon be put to practical application [21].

### Acknowledgements

The research carried out at ATOMKI was supported by the TKP2021-NKTA-42 project financed by the National Research, Development and Innovation Fund of the Ministry for Innovation and Technology, Hungary.

The MGC-20 cyclotron of ATOMKI is a Research Infrastructure of the Cluster of Low Energy Accelerators for Research (CLEAR) of the EURO-LABS project. I.M. Kadenko was supported by the Transnational Access of the CLEAR EURO-LABS project. The EURO-LABS project has received funding from the European Union's Horizon Europe Research and Innovation programme under Grant Agreement No. 101057511.